\documentstyle[aps,preprint]{revtex}
\begin{document}
\draft
\tightenlines
\preprint{TUIMP-TH-96/82}
\title{ Supersymmetric Electroweak 
Corrections to Top Quark Production at LHC }
\author{Hong-Yi Zhou} 
\address{CCAST (World Laboratory), \hspace{.2cm}
      P. O.\hspace{0.2cm}  Box 8730, Beijing 100080, P.R. China,\\
Institute of Modern Physics and Department of Physics,\\
Tsinghua University, Beijing 100084, P.R. China$^*$ \\}

\author{Chong-Sheng Li}
\address{CCAST (World Laboratory), \hspace{.2cm}
P. O.\hspace{0.2cm}  Box 8730, Beijing 100080, P.R. China,\\
Physics Department, Peking University, Beijing 100871, P.R. China$^*$}

\date{\today}
\maketitle
\begin{abstract}
The $\alpha m_t^2/m_W^2$ order  
supersymmetric electroweak corrections arising from loops 
of chargino, neutralino, and squark  to top quark pair 
production by $gg$ fusion at LHC  are calculated in the 
minimal supersymmetric model. We found that the 
corrections amount about a few percent.
\end{abstract}


\vfill

\newpage
\begin{flushleft} 
{\bf I. Introduction} 
\end{flushleft}

The top quark has been found experimentally 
by the CDF and D0 Collaborations with the mass and production
cross section $m_t=176\pm 8(stat)\pm 10(syst)$ GeV~~~ 
$\sigma=6.8^{+3.6}_{-2.4} pb$,
~and $m_t=199^{+19}_{-20}(stat)\pm 22(syst)$ GeV~~~
$\sigma=6.4\pm 2.2 pb$, respectively \cite{CDFD0}. 
Although this measured mass is close to the central value predicted 
by the best fit of the Standard Model (SM ) to the latest LEP data,
the central value of the cross section is somewhat larger than the 
Standard Model prediction $\sigma_{t\bar t}=5.52^{+0.07}_{-0.45} $ pb 
for  $m_t=175$ GeV at $\sqrt{s}=1.8$ TeV $p\bar p$ collider 
in which the effects of multiple soft-gluon emissions have been 
properly resummed\cite{BERGER}.  In addition, 
there are still a number of unsolved theoretical problems in the SM.
New physics beyond the SM are still possible. 
Among various models of new physics so far considered, 
supersymmetry (SUSY) is a promising one at present. 
The simplest and interesting SUSY model is the minimal supersymmetric 
extension of the standard model (MSSM) \cite{MSSM}. 
At the future multi-TeV proton colliders such 
as the CERN Large Hadron Collider (LHC), $t{\bar t}$
production will be enormously larger than the Tevatron rates 
and  the accuracy with which the top quark production cross section
can be measured will be much better (the uncertainty is about 5\% at LHC
\cite{TOPPHYSICS}). 
Thus theoretical calculations of the radiative corrections to the 
production of the top quark at those colliders are of importance. 

QCD corrections to $O(\alpha_s^3)$ and electroweak one-loop corrections   
in the SM to $t\bar t$ production in hadron colliders are 
carried out in Ref.\cite{BERGER}\cite{SMQCD}  and Ref.\cite{SMEW}. 
Yukawa corrections to $t\bar t$ production 
at the Fermilab Tevatron and LHC in two-Higgs-doublet models  
are  calculated in \cite{MSSMYK}\cite{MSSMYKLHC}. 
In the MSSM, electroweak corrections from 
chargino, neutralino and squark to the top pair production via 
$q{\bar q}$ annihilation at the Fermilab Tevatron are calculated 
in Ref.\cite{MSSMEW}. 
Recent calculation of the supersymmetric QCD corrections to the top 
quark pair production at the Tevatron shows that they increase the 
cross section by about 20\%\cite{MSSMQCD}\cite{kim}. 
But this is still within the experimental uncertainty 30\%. 
At LHC the main production mechanism of top quark pair is the gluon-gluon 
fusion process $gg\rightarrow t\bar t$. In this paper we investigate 
the electroweak  corrections of order $ \alpha m_t^2/m_W^2$ 
arising from chargino, neutralino and squark to the top quark 
production by the process $gg\rightarrow t{\bar t}$ at LHC.  
The formalism of the calculation of the corrections to the matrix 
elements will be given in Sec.II. In Sec III, we present our 
numerical examples and discussions of the 
corrections to the cross sections in the MSSM.  

\begin{flushleft} 
{\bf II. Formalism} 
\end{flushleft}

The tree-level Feynman diagrams and the relevant supersymmetric electroweak  
corrections to $gg\rightarrow t{\bar t}$ are shown in Fig.1 (u-channel
diagrams of (b) and (e)--(h) are not explicitly shown) in which the 
dashed lines in the loop represent the squark $\tilde{t}_i$ or 
$\tilde{b}_i~(i=1,2)$ and the solid lines  represent neutralinos 
or charginos, respectively.   

The supersymmetric partner of left- and right-handed massive 
quarks mix \cite{TTMX}. 
The mass eigenstates $\tilde{q}_1$ and
$\tilde{q}_2$ are related to the current eigenstates $\tilde{q}_L$ 
and $\tilde{q}_R$ 
\begin{equation}
\tilde{q} _1 = \tilde{q} _L \cos \theta _q + \tilde{q} _R \sin \theta 
_q,~~~\tilde{q} _2 = -\tilde{q} _L \sin \theta _q + \tilde{q} _R \cos 
\theta _q 
\end{equation}  

The mixing angle $\theta_t$ and the masses $m_{\tilde{t}_1}$,
$m_{\tilde{t}_2}$ can be calculated by diagonalizing the following mass 
matrix
\begin{eqnarray}
\label{eqnum2}
& &  M^2_{\tilde{t}} =\left(
           \begin{array} {ll}
             M^2_{\tilde{t}_L} & m_tm_{LR} \\
             m_tm_{LR} & M^2_{\tilde{t}_R} 
           \end{array}
         \right)    \nonumber\\
& & M^2_{\tilde{t}_L}=m^2_{\tilde{t}_L} + m^2_t+(\frac{1}{2}
  -\frac{2}{3}\sin^2\theta_W)\cos(2\beta)m_Z^2\nonumber\\
& &  M^2_{\tilde{t}_R}=m^2_{\tilde{t}_R} + m^2_t
  +\frac{2}{3}\sin^2\theta_W\cos(2\beta)m_Z^2\nonumber\\
& & m_{LR}=\mu\cot\beta+A_t
\end{eqnarray}   
where $m^2_{\tilde{t}_L},~m^2_{\tilde{t}_R}$ are the soft SUSY-breaking 
mass terms of left- and right-handed stops,  $\mu$ is the coefficient 
of the  $H_1~H_2$ mixing term in the superpotential, $A_t$    
is the parameter describing the strength of soft SUSY-breaking trilinear 
scalar interaction $\tilde{t}_L\tilde{t}_R H_2$, 
$\tan\beta=v_2/v_1$ is the ratio of the vacuum expectation values 
of the two Higgs doublets. 

From Eq.(\ref{eqnum2}), we can get the expressions for 
$m^2_{\tilde{t}_{1,2}}$ and $\theta_t$ : 
\begin{eqnarray}  
& & m^2_{\tilde{t}_{1,2}}=\frac{1}{2}\left[ 
M^2_{\tilde{t}_{L}}+M^2_{\tilde{t}_{R}}\mp\sqrt{(M^2_{\tilde{t}_{L}}
-M^2_{\tilde{t}_{R}})^2+4m_t^2m_{LR}^2}\right]\\
& &\tan\theta_t=\frac{m^2_{\tilde{t}_1}-M^2_{\tilde{t}_{L}}}{m_tm_{LR}}
\end{eqnarray}

For the sbottoms, we neglect the mixing between the left- and right-
handed sbottoms($\theta_b=0$) and have
\begin{eqnarray}  
m^2_{\tilde{b}_{1,2}}=m^2_{\tilde{t}_{L},\tilde{b}_{R}} 
+ m^2_b\pm(T^3_{L,R}-Q_b
\sin\theta_W)\cos(2\beta)m^2_Z, 
\end{eqnarray}
where $T^3_{L,R}=-\frac{1}{2},~0$, $Q_b=-\frac{1}{3}$ and 
$m^2_{\tilde{t}_{L},\tilde{b}_{R}}$ are the soft SUSY-breaking 
mass terms for left- and right-handed sbottoms.

In the presence of squark mixing,the squark-quark-neutralino and 
squark-quark-chargino 
interaction Lagrangian of order $gm_t/m_W$ is given by\\
\begin{eqnarray}
 L_{\tilde{\chi} \tilde{q} \bar{q}} & = & -\frac{gm_t}{\sqrt{2}m_W\sin\beta} 
\sum\limits_{j=1}^{4}\bar{t}[(a_{\tilde{t}_1j}-b_{\tilde{t}_1j}
\gamma_5)\tilde{t}_{1}+(a_{\tilde{t}_2j}-b_{\tilde{t}_2j}\gamma_5)
\tilde{t}_{2}]\tilde{\chi}^0_j                \nonumber \\
& & +\frac{gm_t}{\sqrt{2}m_W\sin\beta} 
\sum\limits_{j=1}^{2}\bar{t}[(a_{\tilde{b}_1j}-b_{\tilde{b}_1j}
\gamma_5)\tilde{b}_{1}+(a_{\tilde{b}_2j}-b_{\tilde{b}_2j}\gamma_5)
\tilde{b}_{2}]\tilde{\chi}^+_j+H.C.\;,
\end{eqnarray}          
where $g$ is the SU(2) coupling constant , and 
$a_{\tilde{t}_1j},~b_{\tilde{t}_1j},~a_{\tilde{t}_2j},
~b_{\tilde{t}_2j},~a_{\tilde{b}_1j},~b_{\tilde{b}_1j},
~a_{\tilde{b}_2j},~b_{\tilde{b}_2j} $ are given by
\begin{eqnarray}
\label{ab}
& & a_{\tilde{t}_1j}=\frac{1}{2}(N_{j4}^\ast\cos\theta_t+N_{j4}\sin\theta_t), 
~~b_{\tilde{t}_1j}=\frac{1}{2}(N_{j4}^\ast\cos\theta_t-N_{j4}\sin\theta_t),
\nonumber\\
& & a_{\tilde{t}_2j}=\frac{1}{2}(-N_{j4}^\ast\sin\theta_t+N_{j4}\cos\theta_t), 
~~ b_{\tilde{t}_2j}=\frac{1}{2}(-N_{j4}^\ast\sin\theta_t-N_{j4}\cos\theta_t),
\nonumber \\
& & a_{\tilde{b}_1j}=b_{\tilde{b}_1j}=\frac{1}{2}V_{j2}^\ast\cos\theta_b,
~~ a_{\tilde{b}_2j}=b_{\tilde{b}_2j}=-\frac{1}{2}V_{j2}^\ast\sin\theta_b
\end{eqnarray}  
$V_{j2}$  are the elements of $2\times 2 $ matrix V and $N_{j4}$ are  
the elements of $4\times 4 $ matrix N (see the Appendix).
 
At the tree level, the S-matrix element is composed of three different 
production channels(s-,t-,u-channel) as follows:

\begin{eqnarray}                                                
 M_0^{s} &= & -ig_s^2(if_{abc}T^c)_{ji}{\bar u}(p_2)
 \rlap/\Gamma v(p_1)/\hat{s}\nonumber\\
&=&-i(T^aT^b-T^bT^a)_{ji}M_0^{s\prime},  \\
 M_0^t& = & -ig_s^2(T^bT^a)_{ji}{\bar u}(p_2)\rlap/\epsilon_4
 (\rlap/q+m_t)\rlap/\epsilon_3 v(p_1)/(\hat{t}-m_t^2)\nonumber\\
& =& -i(T^bT^a)_{ji}M_0^{t\prime},\\
 M_0^u& = & M_0^t(p_3\leftrightarrow p_4,\;\;T^a\leftrightarrow T^b,\;\;
 \hat{t}\rightarrow \hat{u} )\nonumber\\
&=&-i(T^aT^b)_{ji}M_0^{u\prime},
\end{eqnarray}        
where  $q=p_2-p_4$, $\epsilon_4^\mu=\epsilon^\mu(p_4),~~
\epsilon_3^\mu=\epsilon^\mu(p_3)$, 
$\Gamma^\mu$ is given in the Appendix. Instead of 
calculating the square of the amplitudes explicitly,  we calculate  
the helicity amplitudes numerically by using the method of Ref. 
\cite{ZEPPEN}. This method greatly simplifies our calculations.   

The $O(\alpha m_t^2/m_W^2)$ SUSY electroweak corrections to 
$gg\rightarrow t{\bar t}$ are shown in Fig.1 (c)--Fig.1 (h). 
The sum of them is QCD gauge invariant without the strong coupling
constant renormalization . 
In our calculation, we use dimensional regularization to regulate 
the ultraviolet divergences and adopt the on-mass-shell 
renormalization scheme. We also discard the terms proportional 
to $\gamma_5$. 

We only give the explicitly results of the s- and t-channel contributions to 
the SUSY electroweak corrections. The u-channel results can be obtained by the 
following substitutions:
\begin{eqnarray}                               
& & p_3\leftrightarrow p_4,\;\;
T^a\leftrightarrow T^b,\;\;
\hat{t}\leftrightarrow\hat{u}.
\end{eqnarray}        

Fig.1(c) lead to the s-channel vertex correction $\delta M^{s1}$:
\begin{eqnarray}                                   
\delta M^{s1}& =& -ig_s^2 (if_{abc}T^c)_{ji}
{\bar u}(p_2)
 (F_0^{s1}+F_1^{s1}\rlap/\Gamma+\rlap/F_6^{s1} )v(p_1)/\hat{s}\nonumber\\
& =&-i(T^aT^b-T^bT^a)\delta M^{s1\prime} 
\end{eqnarray}                  

Fig.1(d) gives $\delta M^{s2}$:
\begin{eqnarray}                                 
 \delta M^{s2}& =&-ig_s^2(T^aT^b+T^bT^a)_{ji}
{\bar u}(p_2)F_0^{s2}v(p_1)\nonumber\\
& = &-i(T^bT^a+T^aT^b)_{ji}
\delta M^{s2\prime}
\end{eqnarray}

The top quark self-energy $\delta M^{self,t}$ of Fig.1(e) is:  
\begin{eqnarray}                                
\delta M^{self,t}&=& -ig_s^2(T^bT^a)_{ji}
  {\bar u}(p_2)\rlap/\epsilon_4(\rlap/q+m_t)\nonumber\\
& &[F_1^{self,t}+F_2^{self,t}\rlap/q]
  (\rlap/q+m_t)\rlap/  \epsilon_3 v(p_1)/(\hat{t}-m_t^2)\nonumber\\
& =&-i(T^bT^a)_{ji}\delta M^{self,t\prime}
\end{eqnarray}

Vertex correction $\delta M^{v1,t}$ of Fig.1(f) is:
\begin{eqnarray}                                 
\delta M^{v1,t}&=&  \displaystyle -ig_s^2(T^bT^a)_{ji}
{\bar u}(p_2)(F_0^{v1,t}+F_1^{v1,t}\rlap/\epsilon_4+\rlap/F_6^{v1,t})
(\rlap/q+m_t)\rlap/\epsilon_3 v(p_1)/(\hat{t}-m_t^2)\nonumber\\
&=&-i(T^bT^a)_{ji} \delta M^{v1,t\prime}
\end{eqnarray}

Vertex correction $\delta M^{v2,t}$ of Fig.1(g) is:
\begin{eqnarray}                                 
\delta M^{v2,t}&=& \displaystyle -ig_s^2(T^bT^a)_{ji}
{\bar u}(p_2)\rlap/\epsilon_4(\rlap/q+m_t)
(F_0^{v2,t}+F_1^{v2,t}\rlap/\epsilon_3+\rlap/F_6^{v2,t})
  v(p_1)/(\hat{t}-m_t^2)\nonumber\\
& =&-i(T^bT^a)_{ji}\delta M^{v2,t\prime}
\end{eqnarray}

$\delta M^{box,t}$ of the box diagram Fig.1(h) is:
\begin{eqnarray}                      
\delta M^{box,t}& =&
\displaystyle -ig_s^2(T^bT^a)_{ji}
{\bar u}(p_2)[F_0^{b,t}+\rlap/F_3^{b,t}]v(p_1)\nonumber \\
& =&-i(T^bT^a)_{ji}\delta M^{box,t\prime}
\end{eqnarray}

The total amplitude can be written as:
\begin{eqnarray} 
& & M_{ji}=-i[(M_0^+ +\delta M^+)  O^{(+)}_{ji}
+(M_0^- +\delta M^-)O^{(-)}_{ji}~],                    
\end{eqnarray} 
where
\begin{eqnarray*} 
 & & O^{(+)}=\frac{T^bT^a+T^aT^b}{2}\;,\;\; O^{(-)}=\frac{T^bT^a-T^aT^b}{2}\\
 & & M_0^+=M_0^{t\prime}+M_0^{u\prime}\;,\;\;  M_0^-=M_0^{t\prime}-M_0^{u\prime}
-2M_0^{s\prime},
\end{eqnarray*}
\begin{eqnarray} 
\delta M^+&=& 2\delta M^{s2\prime}+\delta M^{self,t\prime}+\delta 
M^{self,u\prime}+\delta M^{v1,t\prime}+\delta M^{v1,u\prime}\nonumber\\
& & +\delta M^{v2,t\prime}+\delta M^{v2,u\prime}+\delta M^{box,t\prime}
+\delta M^{box,u\prime}\nonumber\\
\delta M^-&=&-2\delta M^{s1\prime}+\delta M^{self,t\prime}-\delta 
M^{self,u\prime}+\delta M^{v1,t\prime}-\delta M^{v1,u\prime}\nonumber\\
& &+\delta M^{v2,t\prime}-\delta M^{v2,u\prime}  
+\delta M^{box,t\prime}-\delta M^{box,u\prime}
\end{eqnarray}                   

The color sum of the corrected amplitude square is:
\begin{eqnarray}  
\sum\limits_{color}|M|^2&=&\frac{7}{3}|M_0^+|^2+3|M_0^-|^2\nonumber\\
& &+\frac{14}{3}Re(M_0^+\delta M^{+\dagger})+6Re(M_0^-\delta M^{-\dagger})
\end{eqnarray}
The spin sum as well as phase-space integration and parton distribution 
convolution are done by the  VEGAS program. 
The correction cross section $\Delta \sigma$ is defined  as
\begin{equation}
\Delta \sigma=\sigma-\sigma_0,       
\end{equation}        
where $\sigma$ is the cross section given by $|M|^2$ and $\sigma_0$ 
is the tree level QCD cross section given by $|M_0^+|^2$ and $|M_0^-|^2$ .

\begin{flushleft} 
{\bf III. Numerical Examples and Discussions} 
\end{flushleft}

The production cross section is obtained by convoluting the partonic
cross section with certain parton distribution. The relative correction 
is not sensitive to parton distribution.  
In this paper, we take the
Martin-Roberts-Stirling (MRS) parton distribution set A$^\prime$ \cite{MRSA}
with $Q^2=\hat{s}$.
The following parameters are used in our calculation:

\begin{eqnarray}                                  
& & \sqrt{s}=14~TeV,\;\;m_t=176~GeV,~m_b=4.9 ~GeV,\nonumber\\
& & m_W=80.22~GeV,~m_Z=91.175~GeV,~\alpha=\frac{1}{128.8} .  
\end{eqnarray} 

\noindent
Care must be taken in the calculation of the form factors expressed in
terms of the standard loop integrals defined in Ref. \cite{VELTMAN}.
As has been discussed in Ref.\cite{DENNER}, the formulae for the form 
factors given in terms of the tensor loop integrals will be ill-defined when 
the scattering is forwards or backwards wherein the Gram determinants of some 
matrices vanish and thus their inverses do not exist. This problem 
can be avoided by taking the kinematic  cuts on the rapidity $y$
and the transverse momentum $p_T$. In this paper, we take
\begin{equation}                                                     
|y|<2.5,\;\;p_T>20\;GeV.\\
\end{equation}
The cuts will also increase the relative corrections\cite{SMEW}.

We first checked the QCD gauge invariance by the substitution 
$p_4\rightarrow\epsilon_4$ and $p_3\rightarrow\epsilon_3$ and 
find that $\delta M^+,~\delta M^-$ are a few order of magnitudes 
smaller.  

In the calculation  of the chargino and neutralino masses, 
we fix $M=200~GeV$, $\mu=-100~GeV$ and use the relation $M^\prime 
=\frac{5g^{\prime2}}{3g^2}M$(see the Appendix). We also assume 
$m_{\tilde{t}_L}=m_{\tilde{t}_R}=m_{\tilde{b}_R}=m_{\tilde{q}}$. 

The relative correction to the hadronic cross section as a function 
of the squark mass parameter $m_{\tilde{q}}$ with $\tan\beta=1$ 
and $m_{LR}=0$(corresponding to non-mixing case) is presented in Fig.2.  
For $\tan\beta=1$, the chargino masses $m_{\tilde{\chi}^+_{j}}=
(220,120)~GeV$ and the neutralino masses  $m_{\tilde{\chi}^0_{j}}=
(105,221,128,100)~GeV$. The correction is always negative.  
For $m_{\tilde{q}}<150~GeV$, the correction is very sensitive 
to $m_{\tilde{q}}$. A sharp dip at about $m_{\tilde{q}}=56~GeV$ 
is due to the singularity of the top quark wave function renormalization 
constant at the threshold point $m_t=m_{\tilde{b}_1}+m_{\tilde{\chi}^+_{2}}$
(note that for $\tan\beta=1$, $m_{\tilde{b}_1}=m_{\tilde{q}}$). 
This singularity will disappear if the finite widths of the 
top quark and the charginos are taken into account. 
The correction exceeds -5\% only in a small region near the dip.   
The correction approaches to zero at large $m_{\tilde{q}}$ 
which shows the decoupling behaviour.  

Fig.3 shows the dependence of the relative correction to the hadronic 
cross section on the stop mixing parameter $m_{LR}$. We set 
$m_{\tilde{q}}=100~GeV$ and $\tan\beta=1$.  $m_{LR}$ affects the mass 
splitting and mixing angle of $\tilde{t}_1$ and $\tilde{t}_2$. 
The mass splitting increases as $m_{LR}$ increases.  
We fix $\tilde{t}_1$ to be the light one(cf. Eq.(3)). Therefore, 
$\theta_t=\frac{\pi}{4}$ for $m_{LR}<0$, 
$\theta_t=-\frac{\pi}{4}$ for $m_{LR}>0$. The mixing angle 
causes the asymmetry of the relative correction between $m_{LR}>0$ 
and $m_{LR}<0$ although the mass splitting of $\tilde{t}_1$ and 
$\tilde{t}_2$ is symmetry between $m_{LR}>0$ and $m_{LR}<0$. 
The dip at about $m_{LR}=-200~GeV$ is due to the threshold effect 
at $m_t=m_{\tilde{t}_1}+m_{\tilde{\chi}^0_j}$. No dip is found 
at $m_{LR}=200~GeV$ because when $m_{LR}>0$, $\theta_t=-\pi/4$, 
the $t\tilde{t}_1\tilde{\chi}^0_j$ coupling is proportional to 
$\gamma_5$ ($a_{ij}=0$, $b_{ij}\neq 0$). 
From the expression of  the top quark renormalization  constant 
(see the Appendix), one can see that the singularities of $G_0$ and $G_1$ 
cancel with each other when $a_{ij}=0$, $b_{ij}\neq 0$. 

In Fig.4, we present the $\tan\beta$ dependence of the relative correction 
to the hadronic cross section at given $m_{\tilde{q}}=100$ 
GeV and $m_{LR}=100$ GeV. $\tan\beta$ slightly affects the stop mass splitting and $m_{\tilde{b}_1}$. 
The factor $1/\sin^2\beta$ in the coupling constant leads to the rapid 
increase of the correction in the range $\tan\beta<1$. But 
the increase  is somewhat more quickly than $1/\sin^2\beta$ because 
$m_{\tilde{b}_1}$ decreases as $\beta$ decreases. 
  
From Fig.2--4, we see that only for $\tan\beta<1$ and a small region 
near the threshold $t\to \tilde{b}_1 \tilde{\chi}_j^+$ and 
$t\to \tilde{t}_1 \tilde{\chi}_j^0$ the correction may exceed 
$-5\%$. Otherwise, the correction amounts only a few percent smaller 
than $-5\%$. Therefore, we conclude that 
the supersymmetric electroweak 
corrections of order $\alpha m_t^2/m_W^2$ to top quark pair 
production at LHC are potentially observable for $\tan\beta<1$ and 
small parameter region  near the threshold 
$t\to \tilde{b}_1 \tilde{\chi}_j^+$, $t\to \tilde{t}_1 \tilde{\chi}_j^0$ .

\begin{center}{\bf  ACKNOWLEDGMENTS} \end{center}

This work is supported in part by the National Natural Science Foundation of
China, the Fundamental Research Foundation of Tsinghua University and 
a grant from the state commission of Science and Technology of China.
 
\newpage
\begin{center}
{\bf Appendix }
\end{center}

We give here the form factors for the matrix element appeared in the text.  
They are written in terms of the conventional one-, two-, three- and 
four-point scalar loop integrals defined in Ref.\cite{VELTMAN}.  

\begin{eqnarray*}
F_0^{s1}&= & -\sum\limits_{j}\sum\limits_{i=\tilde{t}_1,\tilde{t}_2,
\tilde{b}_1,\tilde{b}_2}CPL [m_j(a_{ij}^2-b_{ij}^2)((p2-p1)\cdot\Gamma C_0\\
& & -2C^{10}\cdot\Gamma)](-p_2,k,m_j,m_i,m_i)\\
F_1^{s1}&= & \sum\limits_{j}\sum\limits_{i=\tilde{t}_1,\tilde{t}_2,
\tilde{b}_1,\tilde{b}_2}CPL [\delta Z_{ij}^v]\\
F_6^{s1\mu}&= & -\sum\limits_{j}\sum\limits_{i=\tilde{t}_1,\tilde{t}_2,
\tilde{b}_1,\tilde{b}_2}CPL [(a_{ij}^2+b_{ij}^2)((p2-p1)\cdot\Gamma C^{10\mu}\\
& & -2C^{21\mu}(\Gamma))](-p_2,k,m_j,m_i,m_i)\\
  & & \\
F_0^{s2}&= & -\sum\limits_{j}\sum\limits_{i=\tilde{t}_1,\tilde{t}_2,
\tilde{b}_1,\tilde{b}_2}CPL [m_t(a_{ij}^2+b_{ij}^2)C_{11}
-m_j(a_{ij}^2-b_{ij}^2)C_{0}](-p_2,k,m_j,m_i,m_i)\\
  & & \\
F_1^{self,t}& =& \sum\limits_{j}\sum\limits_{i=\tilde{t}_1,\tilde{t}_2,
\tilde{b}_1,\tilde{b}_2}CPL [-m_j(a_{ij}^2-b_{ij}^2)B_0+\delta m_{ij}
+m_t\delta Z_{ij}^v](\hat{t},m^2_j,m^2_i)/(\hat{t}-m^2_t)\\
F_2^{self,t}& =& \sum\limits_{j}\sum\limits_{i=\tilde{t}_1,\tilde{t}_2,
\tilde{b}_1,\tilde{b}_2}CPL [(a_{ij}^2+b_{ij}^2)B_1
-\delta Z_{ij}^v](\hat{t},m^2_j,m^2_i)/(\hat{t}-m_t^2)\\
  & &\\
F_0^{v1,t}& =& -\sum\limits_{j}\sum\limits_{i=\tilde{t}_1,\tilde{t}_2,
\tilde{b}_1,\tilde{b}_2}CPL [2m_j(a_{ij}^2-b_{ij}^2)(p_2\cdot \epsilon_4C_0
-C^{10}\cdot\epsilon_4)](-p_2,p_4,m_j,m_i,m_i)\\
F_1^{v1,t}&= & \sum\limits_{j}\sum\limits_{i=\tilde{t}_1,\tilde{t}_2,
\tilde{b}_1,\tilde{b}_2}CPL [\delta Z_{ij}^v]\\
F_6^{v1,t\;\mu}&= & -\sum\limits_{j}\sum\limits_{i=\tilde{t}_1,\tilde{t}_2,
\tilde{b}_1,\tilde{b}_2}CPL [2(a_{ij}^2+b_{ij}^2)(p_2\cdot\epsilon_4 
C^{10\mu}\\
& & -C^{21\mu}(\epsilon_4))](-p_2,p_4,m_j,m_i,m_i)\\
 & &\\
F_0^{v2,t}& =& -\sum\limits_{j}\sum\limits_{i=\tilde{t}_1,\tilde{t}_2,
\tilde{b}_1,\tilde{b}_2}CPL [2m_j(a_{ij}^2-b_{ij}^2)(-p_1\cdot \epsilon_3C_0\\
& & -C^{10}\cdot\epsilon_3)](p_1,-p_3,m_j,m_i,m_i)\\
F_1^{v2,t}&= & \sum\limits_{j}\sum\limits_{i=\tilde{t}_1,\tilde{t}_2,
\tilde{b}_1,\tilde{b}_2}CPL [\delta Z_{ij}^v]\\
F_6^{v2,t\;\mu}&= & -\sum\limits_{j}\sum\limits_{i=\tilde{t}_1,\tilde{t}_2,
\tilde{b}_1,\tilde{b}_2}CPL [2(a_{ij}^2+b_{ij}^2)(-p_1\cdot\epsilon_3 
C^{10\mu}\\
& & -C^{21\mu}(\epsilon_3))](p_1,-p_3,m_j,m_i,m_i)\\
 & & \\
F_0^{b,t}& =& \sum\limits_{j}\sum\limits_{i=\tilde{t}_1,\tilde{t}_2,
\tilde{b}_1,\tilde{b}_2}CPL 
[4m_j(a_{ij}^2-b_{ij}^2)(p_2\cdot \epsilon_4 p_1\cdot\epsilon_3D_0
-p_1\cdot\epsilon_3D^{10}\cdot\epsilon_4\\
& & +p_2\cdot\epsilon_4D^{10}\cdot\epsilon_3
-\epsilon_3\cdot D^{21}(\epsilon_4))] (-p_2,p_4,p_3,m_j,m_i,m_i,m_i)\\
F_3^{b,t\mu}& = & \sum\limits_{j}\sum\limits_{i=\tilde{t}_1,\tilde{t}_2,
\tilde{b}_1,\tilde{b}_2}CPL 
[4(a_{ij}^2+b_{ij}^2)(p_2\cdot \epsilon_4 p_1\cdot\epsilon_3D^{10\mu}
-p_1\cdot\epsilon_3D^{21\mu}(\epsilon_4)\\
& & +p_2\cdot\epsilon_4D^{21\mu}(\epsilon_3)
-D^{32\mu}(\epsilon_3,\epsilon_4))] (-p_2,p_4,p_3,m_j,m_i,m_i,m_i)
\end{eqnarray*}

In the above $i,~j$ summation, $j=1,2,3,4,~m_j=m_{\tilde{\chi}^0_j}$ 
for $i=\tilde{t}_1,~\tilde{t}_2$ ,
$j=1,2,~m_j=m_{\tilde{\chi}^+_j}$ for $i=\tilde{b}_1,~\tilde{b}_2$. 
As $\theta_b=0$, $\tilde{b}_2$ actually does not contribute to the 
sum(cf. Eq.(\ref{ab})). 
\begin{eqnarray*}
& & k=p_1+p_2=p_3+p_4,\;\;\hat{s}=k^2,\;\;
\hat{t}=q^2=(p_2-p_4)^2,\;\;\hat{u}=(p_2-p_3)^2,\\
& & \Gamma^\mu=(-p_4+p_3)^\mu\epsilon_3\cdot\epsilon_4+(2p_4+p_3)
\cdot\epsilon_3
\epsilon_4^\mu-(2p_3+p_4)\cdot\epsilon_4\epsilon_3^\mu \;,\\
& & CPL=\frac{\alpha m_t^2}{8\pi m_W^2\sin^2\theta_W\sin^2\beta}
\end{eqnarray*} 

\begin{eqnarray*}
& & C^{10\mu}=C^{\mu}\;, C^{21\mu}(p)=p_\nu C^{\mu\nu}\;,
    C^{20}=g_{\mu\nu}C^{\mu\nu}-\displaystyle\frac{1}{2}\;,\\
& & D^{10\mu}=D^{\mu}\;, D^{21\mu}(p)=p_\nu D^{\mu\nu}\;,
    D^{20}=g_{\mu\nu}D^{\mu\nu}\;,\\
& & D^{32\mu}(p,l)=p_\nu l_\alpha D^{\mu\nu\alpha}\;,  
    D^{30\mu}=g_{\nu\alpha}D^{\mu\nu\alpha}\;. 
\end{eqnarray*} 

In our calculation, we calculate the tensor loop integrals 
$C^\mu$, $C^{\mu\nu}$, $D^{\mu}$, $D^{\mu\nu}$ and $D^{\mu\nu\alpha}$ 
numerically  instead of  expanding them explicitly.

The renormalization constants are:\\
\begin{eqnarray*}
& & Z^v_{ij}=[(a_{ij}^2+b_{ij}^2)(B_1+2m_t^2G_1)+2(a_{ij}^2-b_{ij}^2)
m_tm_jG_0](p^2,m^2_j,m^2_i)|_{p^2=m_t^2}\\
& & \delta m_{ij}=[-(a_{ij}^2+b_{ij}^2)m_tB_1+(a_{ij}^2-b_{ij}^2)m_jB_0]
(p^2,m^2_j,m^2_i)|_{p^2=m_t^2}
\end{eqnarray*}
where  $\displaystyle G_0=-\frac{\partial B_0(p^2,m_i^\prime,m_i)}
{\partial p^2}$, 
$\displaystyle G_1=\frac{\partial B_1(p^2,m_i^\prime,m_i)}{\partial p^2}$ .

The neutralino and chargino masses , the $2\times 2$ matrix $V$ and 
the $4\times 4$ matrix $N$ are given by the following relations\cite{MSSM}:
$$
  X =\left(
           \begin{array} {cc}
             M & m_W\sqrt{2}\sin\beta \\
             m_W\sqrt{2}\cos\beta & \mu
           \end{array}
         \right)~,
$$
$$ VX^2V^{-1}=M_D^2 $$

$$
  Y =\left(
           \begin{array} {cccc}
             M^\prime & 0 & -m_Z s_W\cos\beta & m_Z s_W\sin\beta\\
             0 & M & m_Zc_W\cos\beta & -m_Zc_W\sin\beta \\
             -m_Z s_W\cos\beta & m_Zc_W\cos\beta & 0 & -\mu\\
             m_Z s_W\sin\beta & -m_Zc_W\sin\beta & -\mu & 0
           \end{array}
         \right)~,
$$

$$ N Y^2 N^{-1}=N_D^2 \;,\;\;N^\ast Y N^{-1}=N_D~, $$
where $M$, $M^\prime$ are the masses of gauginos corresponding to $SU(2)$
and $U(1)$, respectively. With the grand unification assumption, we have 
the relation $M^\prime=\frac{5}{3}(g^{\prime 2}/g^2)M$. 
$M_D^2$ is a diagonal matrix and the chargino mass squares correspond 
to its diagonal elements.
$N_D$ is a diagonal matrix with non-negative entries which give the masses 
of neutralinos.  The chargino and  neutralino masses depend on the parameters
$M$, $\mu$ and $\tan\beta$.

\newpage

\begin{center}
{\large Figure Captions}
\end{center}

\parindent=0pt

Fig.1 Feynman diagrams of tree level and $O(\alpha m_t^2/m^2_W)$ 
SUSY electroweak correction.  

Fig.2 Relative correction to hadronic cross section versus 
$m_{\tilde{q}}$ with $\tan\beta=1$ and $m_{LR}=0$.

Fig.3 Relative correction to hadronic cross section versus $m_{LR}$ with 
$\tan\beta=1$ and  $m_{\tilde{q}}=100$ GeV.  

Fig.4 Relative correction to hadronic cross section versus  
$\tan\beta$  with  $m_{\tilde{q}}=100$ GeV and $m_{LR}=100$ GeV.   





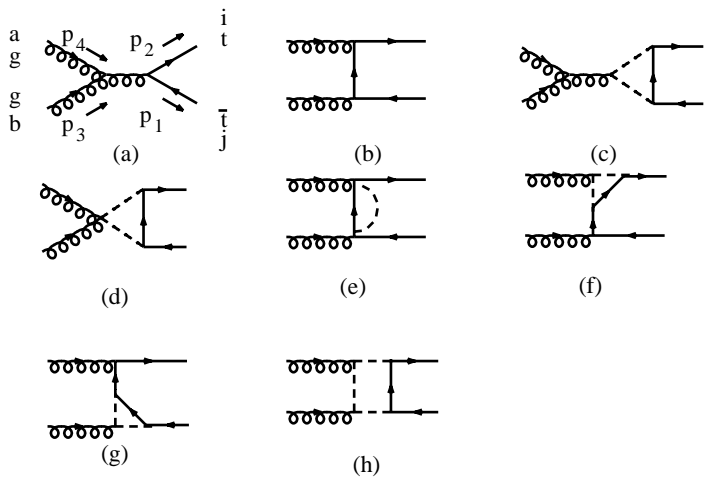
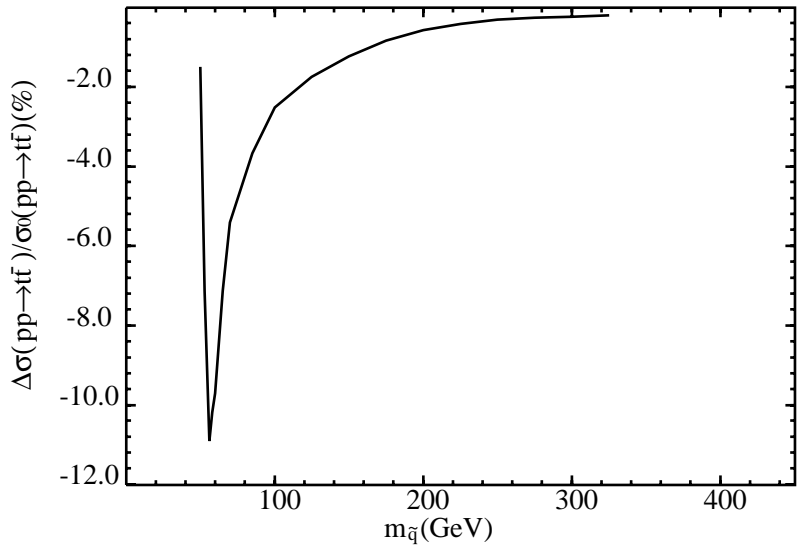
\end{document}